# Inspiration from machine learning on example of optimization of the Bose-Einstein condensate of thulium atoms in a 1064-nm trap

D.A. Kumpilov[1,2], D.A. Pershin [1,3], I.S. Cojocaru[1,3,4], V.A. Khlebnikov[1], I.A. Pyrkh[1,2], A.E. Rudnev[1,2], E.A. Fedotova[1], K.A. Khoruzhii[1], P.A. Aksentsev[1,4], D.V. Gaifutdinov[1], A.K. Zykova[1], V.V. Tsyganok [1], A.V. Akimov[1,3]

[1]*Russian Quantum Center, Bolshoy Boulevard 30, building 1, Skolkovo, 143025, Russia*

[2]*Moscow Institute of Physics and Technology, Institutskii pereulok 9, Dolgoprudny, Moscow Region 141701, Russia*

[3]*PN Lebedev Institute RAS, Leninsky Prospekt 53, Moscow, 119991, Russia*

[4]*Bauman Moscow State Technical University, 2-nd Baumanskaya, 5, Moscow, 105005, Russia*

email: *a.akimov@rqc.ru*

The number of atoms in Bose-Einstein condensate determines the scale of experiments that can be performed, making it crucial for quantum simulations. Optimization of the number of atoms in the condensate is a complex problem which could be efficiently solved using machine learning technique. Nevertheless, this approach usually does not give any insight in the underlying physics. Here we demonstrate possibility to learn physics from the machine learning on an example of condensation of thulium atoms at a 1064-nm dipole trap. Optimization of the number of condensed atoms revealed a saturation, which was explained as limitation imposed by a 3-body recombination process. This limitation was successfully overcome by leveraging Fano-Feshbach resonances.

# I. INTRODUCTION

After the first successful cooling of rubidium [1] and sodium [2] atoms to the Bose-Einstein condensation (BEC), the field of cold atom-based quantum degenerate gases rapidly developed [3–11]. One intriguing application of the condensate is quantum simulations, enabling the understanding of complex materials through modeling in a controllable quantum system [12]. Among many candidates for quantum simulations, lanthanoids hold a special place due to their large magnetic moment in the ground state, facilitating long-range interactions and a significant number of low-field Fano-Feshbach resonances. These resonances allow detailed control of short-range interactions. The thulium atom has a magnetic dipole moment of four Bohr magnetons ($4\mu_B$) in the ground state and a dense non-chaotic set of Fano-Feshbach resonances [13,14]. Recently, the machine learning approach has made it possible to condense thulium atoms into BEC in a 532-nm dipole trap, providing close atom packing [11].

Moreover, following the pioneering work [15] several groups switched to simulations in a quantum gas microscope [16–21]. This approach offers the advantage of individual atom control instead of relying solely on ensemble measurements. This individual control becomes particularly advantageous when addressing issues such as Anderson localization [22–24]. Therefore, establishing a quantum microscope for the thulium atom could prove highly beneficial.

The necessity to optically resolve nearby sites requires a numerical aperture of an objective to be as high as possible [15–21]. Given realistic limitations regarding a housing of the vacuum volume, the size of an individual lattice size was intended to be set to about micron. For applications involving optical lattices with site sizes ranging from 0.5 to several microns, implementing a dipole trap operating at a wavelength around one micron offers distinct advantages.

In this work, cooling of the thulium atom to the BEC in a 1064-nm dipole trap was performed. The cooling was optimized using the Bayesian machine learning technique that has already demonstrated valuable impact on the performance of experiments with BEC production [25–27]. The machine learning optimization was found to experience saturation at a level of about $2 \cdot 10^4$ atoms in the condensate. This saturation made it possible to conclude that there is an uncontrolled parameter responsible for the limitation. Thus, the spectrum of Feshbach resonances of the thulium atom [13,14] was reevaluated, and another magnetic field for the cooling sequence was selected. This adjustment indeed enabled the optimizer to increase the number of atoms in the condensate.

## II. EXPERIMENTAL SETUP

The experimental setup closely followed the configuration presented in [11]. The Zeeman slower and 2D optical molasses precooled atoms, operate at the strong transition $4f^{13}(^2F^0)6s^2 \rightarrow 4f^{12}(^3H_5)5d_{3/2}6s^2$ with a wavelength of 410.6 nm and a natural width of $\Gamma = 2\pi\gamma = 2\pi \cdot 10.5$ MHz. Following the precooling stage, atoms were loaded into the magneto-optical trap (MOT) operating at the weaker transition $4f^{13}(^2F^o)6s^2 \rightarrow 4f^{12}(^3H_6)5d_{5/2}6s^2$ with a wavelength of 530.7 nm and a natural width of $\Gamma = 2\pi\gamma = 2\pi \cdot 345.5$ kHz [28]. The large detuning of the MOT light along with the reduction of its intensity provided the polarization of atoms at the lowest magnetic sublevel $|F = 4; m_F = -4\rangle$ of the ground state [29–34]. The atoms were cooled down to $22.5 \pm 2.5$ µK and then loaded into the optical dipole trap (ODT) (Figure 1A). The ODT was formed by a linearly polarized laser beam ("horizontal beam $P_H$") with a wavelength of 1064 nm focused on the beam waist of $24.0 \pm 0.4$ and $54.2 \pm 0.7$ µm. The beam was scanned using an acousto-optic modulator to increase the geometrical overlap of the ODT and MOT potentials [35]. The second "Vertical beam $P_V$" with a waist of $100 \pm 4$ µm formed the crossed ODT (cODT) potential to increase confinement and collision rate during evaporation. In the cODT, the evaporation was performed via sequential ramps of the power (Figure 1B) of the beams thereby lowering the walls of the trap potential and allowing the higher energy atoms to escape. This stage was the focus of the optimization process. Simple linear parametrization was used, and the parameters specifying the sequence were the endpoints of the linear power ramps.

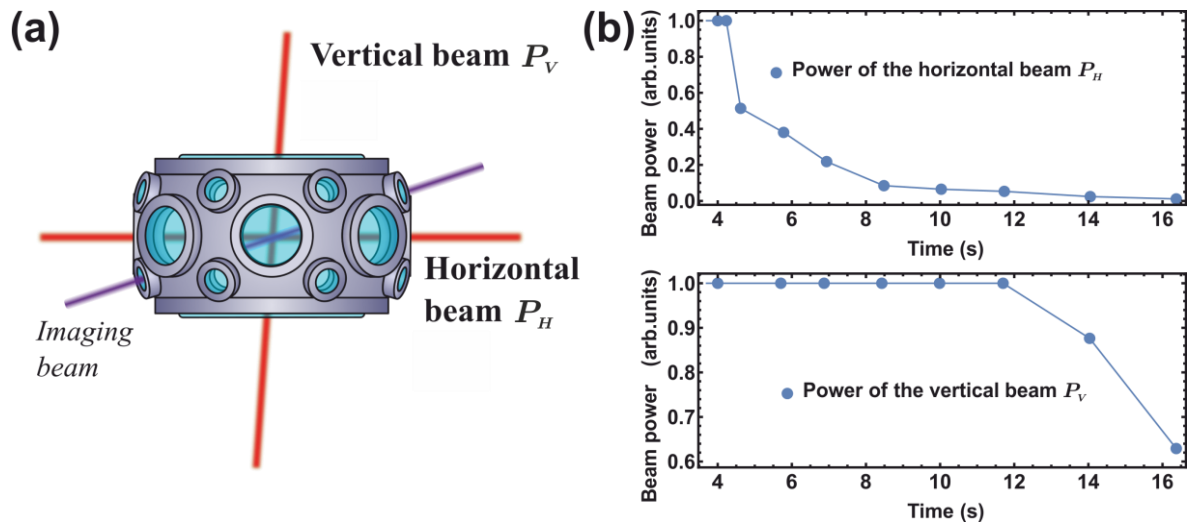

Figure 1. a) Schematic of how the ODT beams are directed to the main UHV cell. Red lines show horizontal and vertical ODT beams; the violet line shows the probe beam for the absorption imaging. b) Evaporation sequence. The blue dots represent the values of beam powers that were set as input parameters for the optimizer.

Imaging of the atoms was achieved through absorption in a probe beam [32]. The atomic cloud, reaching Bose-Einstein condensation, exhibits a bimodal distribution reflected in a two-dimensional absorption image (see Figure 2a). Following a method similar to [33] for determining the number of condensed atoms $N_{BEC}$, a special mask was employed to separate the thermal cloud and the BEC fraction. Data points from the central region of the absorption image overlaying with the mask were excluded, and the remaining data were fitted with a Gaussian distribution (see Figure 2c). The size of the mask was represented by the *s* parameter. The final size parameter *s* was defined as the value when the width $\sigma_x$ of Gaussian distribution ceases to change (Figure 2b). Subsequently, the corresponding Gaussian fit was subtracted from the initial data point-by-point to yield the BEC fraction of the atomic cloud. The BEC fraction data were then fitted with the Thomas-Fermi distribution, providing the BEC size and the number of atoms within it. The initial atomic cloud conforms to the bimodal distribution (see Figure 2d).

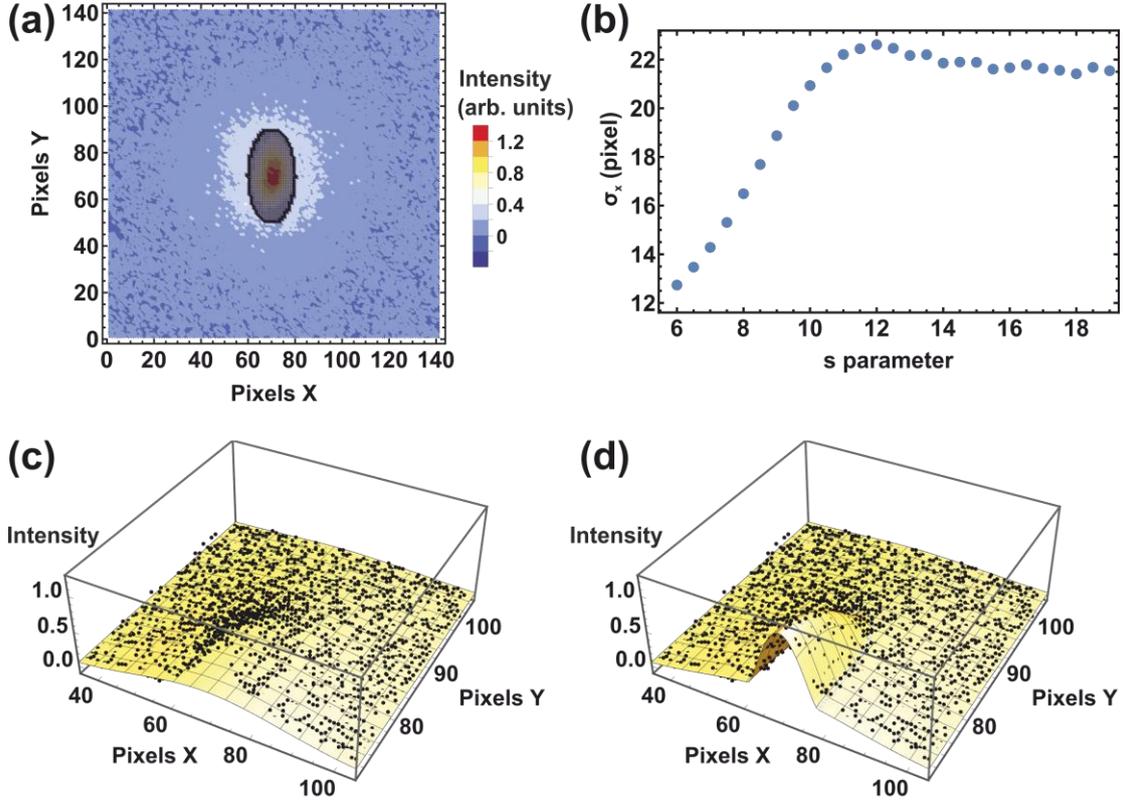

Figure 2. Bimodal fit of the atomic cloud. a) Density plot of the atomic cloud with the presence of BEC. b) The mask used to separate the thermal cloud and the BEC fraction. c) 3D plot of the thermal cloud fit, d) 3D plot of the joint Thomas-Fermi and thermal clouds fits.

## III. OPTIMIZATION PROCEDURE

There are a number of optimization algorithms available; for example, gradient [36], genetic [37], and hybrid [38] algorithms have been used in real-time quantum experiments. Here a feedback-like experimental procedure using a Bayesian machine learning technique based on the Gaussian processes model (for the details see Appendix A) was implemented. Initially, the optimization criterion was the efficiency of evaporation $\gamma$:

$$\gamma = \frac{\ln \frac{D_{PS}}{D_{PS}^{(0)}}}{\ln \frac{N^{(0)}}{N}} \tag{1}$$

where $D_{PS}^{(0)}$ and $D_{PS}$ are the initial and final phase space densities of the atomic cloud, respectively, $N^{(0)}$ and $N$ are the initial and final numbers of atoms, respectively [11].

In the original configuration of the setup, the duration of evaporation step was fixed, and only the intensity of cooling beams was varied. One might question whether this approach is reasonable or whether a change the duration of evaporation step might be beneficial. The answer to this question could be derived from machine learning without necessarily setting the total duration of evaporation step as an optimization parameter. The procedure involves running optimization with the total duration of the evaporation step $t_1$ and then conducting another optimization with $t_2 > t_1$. If the results are the same, setting the total time of the evaporation ramps as an optimization parameter would waste computer resources rather than provide a better result, supposing the dependance of the evaporation efficiency from the duration of evaporation process to be a smooth function.

Contrarily, it was found that, with the cost function (1), the best ramps with a total duration of the evaporation step of 11.8 s were approaching the parameter boundaries, with the PSD reaching 0.034 (the "11.8 s" label in Figure 3b). This clearly underscores that the optimum evaporation for achieving quantum degeneracy should last longer. Consequently, the duration of colling cycle was increased to 14 s. With this adjustment, the best ramps did not reach the boundaries (the "14 s" label in Figure 3b), but the atomic cloud exhibited a PSD of about 0.1, and the duration of evaporation step was still insufficient to achieve condensation. It's worth noting that the horizontal beam power at 11.8 s of the "14 s" evaporation sequence exceeds that of the "11.8 s" sequence, confirming the necessity to increase the total duration of the evaporation step.

The final goal of the experiment is the optimization of the number of atoms in the condensate. Since the phase space density no longer rises once the condensate is achieved, the cost function (1) becomes meaningless. Therefore, to optimize the number of atoms in the condensate, the cost function was modified as:

$$C(\mathbf{X}) = \beta_\gamma \gamma + \beta_{BEC} N_{BEC} \tag{2}$$

where $N_{BEC}$ is the number of atoms in BEC (see Figure 2), $\beta_\gamma$ and $\beta_{BEC}$ are coefficients that can be tailored arbitrarily. This way, the number of atoms is explicitly included in the cost function. It should be noted that cooling efficiency $\gamma$ should also be taken into account since if for some reason condensate was not achieved, the number of atoms in the condensate would be 0. If only the number of atoms in the condensate was the cost function, then all the runs without BEC would be treated equally and the machine would not learn from those. The presence of $\gamma$ helps to solve this issue and to use the experimental runs in the learning process more efficiently.

Considering that the 14 s cooling duration still appears to be too short as mentioned above, two 1.4 s long ramps were added to the evaporation sequence. The optimization routine was performed with the new cost (2) and only 6 parameters – the 3 last powers in the horizontal and vertical beams (the "16.8 Tail" label in Figure 3). The other parameters were set to the best values found in the previous optimization run to ensure high PSD before the BEC formation. The best evaporation sequence produced the BEC with $(2.0 \pm 0.2) \cdot 10^4$ atoms. Note that the horizontal power at the 14 s moment while being a varying parameter had not remarkably changed during this optimization run.

The goal of optimization is to identify the global optimum of an evaporation process [26]. The described iterative procedure likely finds a local optimum, so the optimization was performed with the parameters representing the full evaporative sequence (labeled as "16.8 Full" in Figure 3a). It is noteworthy that the power of the horizontal beam in the "16.8 Full" sequence at 14 s became larger than the one of the "16.8 Tail", while the last two remained unchanged. The number of atoms in the BEC did not change. It underlines the fact that the first ramps of evaporation are not as crucial for the final BEC production.

Finally, the optimization was performed by varying the total duration of the evaporation ramps as a parameter (the "Duration variation" label in Figure 3a), and a slightly shorter 16.3 s sequence with approximately the same powers and the same number of atoms in the BEC was achieved.

Several sequences achieved from optimization procedures are presented in Figure 3a. It is evident that variations in the beam power are not quite trivial. Further details can be found in the APPENDIX A.

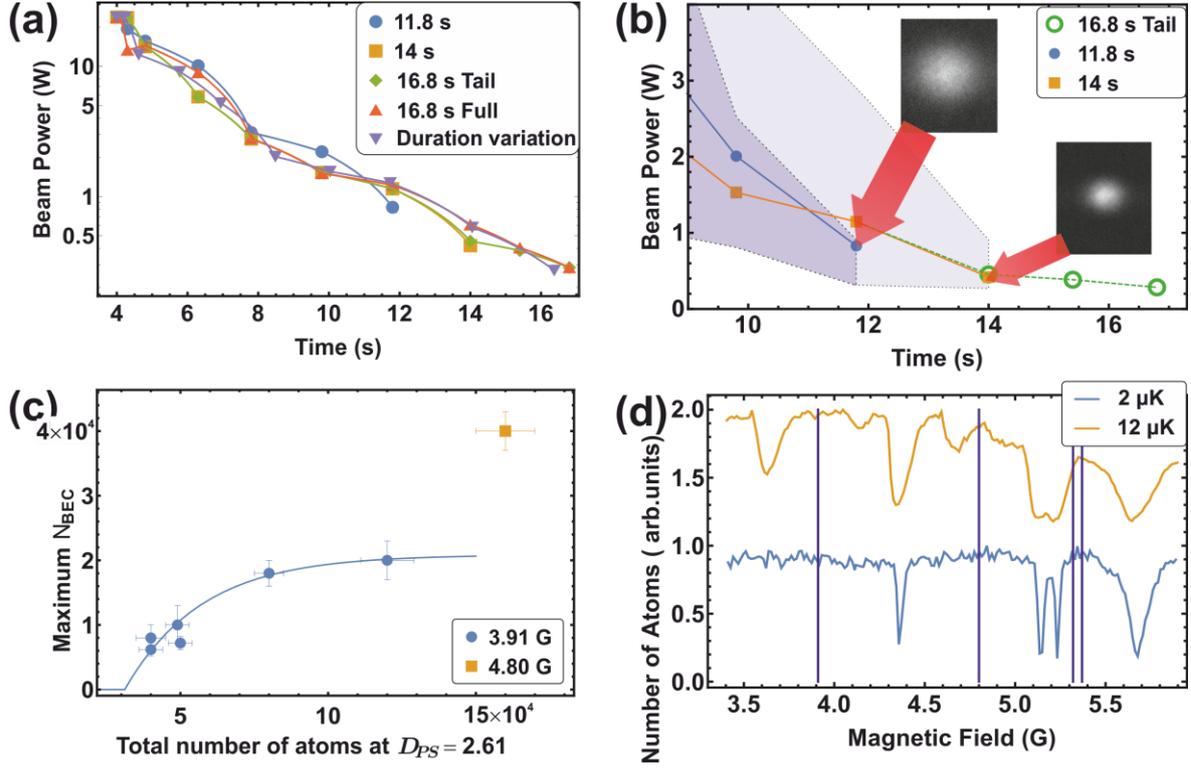

Figure 3. a) Optimized horizontal ODT beam power in the log scale obtained in different optimization runs. b) The process of optimization with the fixed total duration of the evaporation step. The longer the set up total duration, the larger the resulting PSD of the atomic cloud. The shaded regions with dashed and dotted edges represent the boundaries of the search region. The results of optimization are for the duration of the evaporation sequence, the number of atoms, temperature and phase space density are 11.8 s, $6 \cdot 10^5$ atoms, 1.5 µK, 0.034: 14 s, $4 \cdot 10^5$ atoms, 0.8 µK, 0.1; 16.8 s, BEC $2 \cdot 10^4$ atoms, thermal cloud $1.3 \cdot 10^4$ atoms, 0,07 µK; c) The saturation of BEC atoms in one magnetic field and the increment of it in the other. d) Fragment of the Feshbach resonance spectrum, taken from [13]. Vertical lines indicate positions, selected for condensation experiments.

## IV.  UNDERSTANDING OF PHYSICS LIMITING OPTIMIZATION

The optimization technique mentioned above compelled the learner to acquire a large number of atoms with subcritical phase density, leading to a significant increase in the number of atoms with a critical PSD just before BEC formation. However, the number of atoms in the BEC itself did not rise substantially, despite being the optimization criterion. On the contrary, the BEC atom count exhibits a ceiling effect. Specifically, the number of atoms at a critical $PSD = 2.61$ (see chapter V.B at [39]) increases from one experiment to another, but the number of atoms in the condensate remains more or less the same (Figure 3c, label "3.91G"). This observation suggests that there is likely some inelastic process during the evaporation cooling, which becomes strongly enhanced when the BEC starts to form. The dynamics of evaporation cooling were extensively studied [40] in the regime of low phase density, which however does not seem to yield any surprises. In the regime of high PSD, when condensate starts to form, one might expect a decrease in the 3-body recombination cross-section [41] on the one side, but a considerable rise in the density of atoms in the BEC region on the other. Thus, given the saturation observed, one can conclude that 3-body recombination, examined previously for $^{87}$Rb atoms [42,43], is likely to be a limiting factor for evaporation cooling and loading a large number of thulium atoms into the BEC.

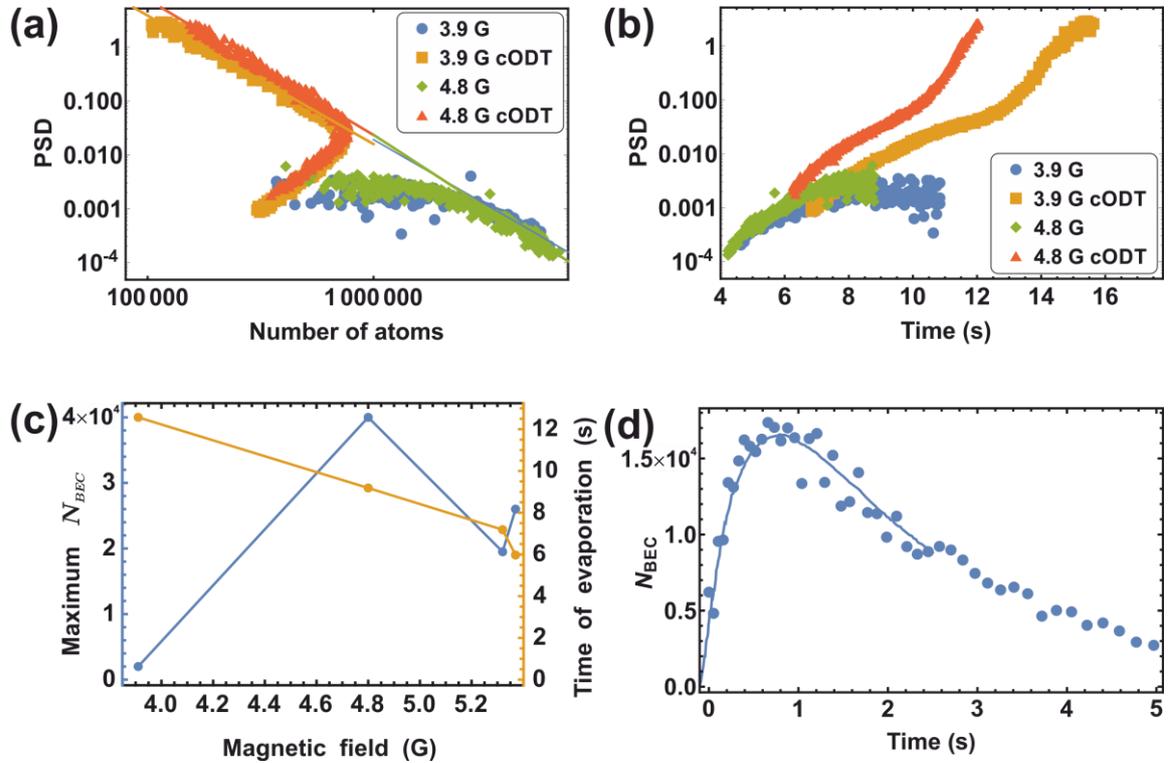

Figure 4. a) Calculated dependence of the PSD versus the number of atoms. PSD is calculated separately for the horizontal and crossed ODT, and values for the both atomic clouds are presented in the intermediate region (see APPENDIX B). b) Phase space density (PSD) of the atomic cloud during the cooling sequence. c) The dependence of the maximum achieved $N_{BEC}$ and the total duration of evaporation step in the ODT on the magnetic field. d) Decay of the thulium BEC and comparison with the theoretical curve from [44] (solid line). The theoretical curve

was rescaled in horizontal and vertical directions by the factors 0.34 and 0.130 respectively.

The 3-body recombination may be strongly enhanced in the middle of Fano-Feshbach resonance. These resonances were studied before for the thulium atom and exhibited strong temperature dependence [13,14]. At higher temperatures, higher-order resonances appear in the spectrum (see Figure 3d). Nevertheless, these resonances at lower temperatures are not completely forbidden, they are just suppressed by a centrifugal barrier, but should still have a finite cross-section. In the regime of large density, the overall probability of high-order Feshbach resonance may become significant and lead to 3-body recombination. It's important to note that this logic had not been confirmed by any calculation and had been a pure guess inspired by the optimization behavior. To test this idea, optimization was performed in several magnetic fields for which there were no known types of Feshbach resonances (see Figure 3d).

Initially, the magnetic field of 4.80 G was selected to perform the BEC optimization. The number of atoms in BEC strongly exceeded the value previously found for 3.91 G (Figure 3c, label "4.8G"). The number of atoms right before quantum degeneracy was also expectedly larger. To understand possible reasons for this, the parameters of the atomic cloud were measured at each point of an evaporation sequence, found by the optimization procedure (Figure 4a, b). The diagram in Figure 4a clearly demonstrates that the efficiency of evaporation was better in the magnetic field of 4.80 G in the crossed ODT and did not significantly differ in the horizontal one.

Two other magnetic fields were set up for BEC optimization. Figure 4c illustrates that both the optimal duration of the evaporation step and the number of atoms in the condensate strongly depend on the magnetic field (see APPENDIX C). The shrinkage of evaporation is likely caused by the dependence of scattering length on the magnetic field, but it may also be related to the 3-body recombination loss channel. However, the maximum number of atoms in BEC does not increase with decreasing the duration of the evaporation step.

Finally, after an intensive literature search, it was discovered that the effect of saturation of the number of atoms in BEC due to 3-body recombination was actually theoretically predicted before for $^{87}$Rb atoms in a magnetic trap [44]. One of the key features of this prediction is the gradual decay of BEC, which is, indeed, observed in the experiment (Figure 4d). To compare with the theoretical prediction, the theoretical curve, presented in [44], was fitted to the experimental data by rescaling the curve horizontally and vertically with the factors 0.34 and 0.130, respectively. Of course, thulium parameters differ from ones for rubidium, but there is no doubt that the general trend is the same. Note that 3-body recombination limiting the number of atoms in BEC has been observed for rubidium and sodium atoms in a magnetic trap [45,46]. In that cases the fast evaporation ramps were applied at the final stages of evaporation to reduce the integral effect of 3-body recombination loss to contrast with our method, relying on changing 3-body recombination rate itself via Feshbach resonance.

## V. CONCLUSION

In summary, Bayesian optimization was used to obtain the BEC of thulium atoms in a 1064-nm optical dipole trap. The implemented optimization procedure made it possible to observe the saturation of the number of atoms in BEC, which was interpreted as 3-body recombination caused saturation. Overcoming this saturation involved analyzing the Feshbach resonance spectrum at a relatively high temperature and selecting another magnetic field for the evaporation procedure. As a result, up to $4 \cdot 10^4$ atoms were condensed into the BEC state thus increasing the number of atoms in the condensate by a factor of 2. The behavior of the BEC decay over time was compared with the previous prediction, made for $^{87}$Rb atoms, indeed attributing the decay to a 3-body recombination.

## ACKNOWLEDGMENTS

We thank Georgy Shlyapnikov for the fruitful discussion of the data presented in this paper. This work was supported by Rosatom in the framework of the Roadmap for Quantum computing (Contract No. 868-1.3-15/15-2021 dated October 5, 2021).

## APPENDIX A  OPTIMIZATION DETAILS

### OPTIMIZATION ALGORIGHM

Given $M$ experimental parameters $\mathbf{X} \in \mathbb{R}^M$ subject to optimization and the measure of performance $C(\mathbf{X})$, the optimization problem can be interpreted as finding the global optimum of a black-box function. This problem is addressed automatically via sequential probing of $C(\mathbf{X})$. In a typical optimization loop, the control computer sends parameters $\mathbf{X}_i$ to the experimental apparatus, receives and evaluates $C(\mathbf{X}_i)$, and decides which parameters $\mathbf{X}_{i+1}$ to probe next. The Bayesian optimization algorithm was applied, which builds a statistical model of $C(\mathbf{X})$ (which is called "cost function" in the context of machine learning) and utilizes it to choose $\mathbf{X}_{i+1}$.

To perform the Bayesian optimization, one should choose a type of model and a strategy to update the model. Following the previous results [11], the most common type of model was exploited – the Gaussian process, which considers $C(\mathbf{X})$ as a stochastic Gaussian process. In other words, at any $\mathbf{X}$, there is a random variable with a gaussian Distribution $p(\mu(\mathbf{X}), \sigma(\mathbf{X}))$ with a mean $\mu(\mathbf{X})$ and a standard deviation $\sigma(\mathbf{X})$. The update strategy is probing the minimum of a biased cost function $a(\mathbf{X}) = \chi_\mu \mu(\mathbf{X}) + \chi_\sigma \sigma(\mathbf{X})$. Different combinations of coefficients $\chi_\mu$ and $\chi_\sigma$ make it possible to perform different types of strategies. For example, if $C(\mathbf{X})$ should be globally minimized, the combination $\chi_\mu = 1, \chi_\sigma = 0$ represents an "optimizer" strategy, and $\chi_\mu = 0, \chi_\sigma = -1$ represents a "scientist" one [37]. To avoid trapping in the local minimum, a balance between these two strategies should be maintained. Therefore, a cycle with 4 $\mathbf{X}$ points was chosen: three points corresponded to the minimum of a biased cost function with the coefficients from the sets $\chi_\mu = \{1, 0, 1\}, \chi_\sigma = \{0, -1, -1\}$ plus one randomly chosen $\mathbf{X}$ point. To build the initial Gaussian model the optimization procedure started with 30 runs with randomly chosen parameters. Then the 4-step cycle mentioned above was repeated, after every of which the updated model was built. Generally, the best cost value was reached in 100-300 iterations (including 30 first random ones) for launches with the number of parameters $M$ ranging from 6 to 16.

## COMMENTS ON BEAM SCANNING

The process of turning off the horizontal beam scanning (refer to the main text) underwent optimization through numerous runs including those described in the main text. All the obtained evaporation sequences indicated that the optimal strategy was turning off the sweeping with a fast 100-ms ramp. Alternative approaches such as employing several long ramps or sweeping the beam throughout the entire evaporation process, resulted in poorer PSD due to a smaller number of atoms compared to the "fast turn off" strategy. This was observed despite the fact that the temperature immediately increased up to 50 µK after turning off. That is why in the subsequent optimization runs, the starting point was fixed at the moment when the sweeping was turned off.

## TYPES OF OPTIMIZATIONS

The best evaporation ramps for different optimization types in the 3.91 G magnetic field are presented in Figure 5a-c. Note that the "Duration variation" sequence, achieved with the total duration of the ramps as a parameter, is slightly shorter than set earlier. Besides, the aspect ratio of the BEC is presented in Figure 1d.

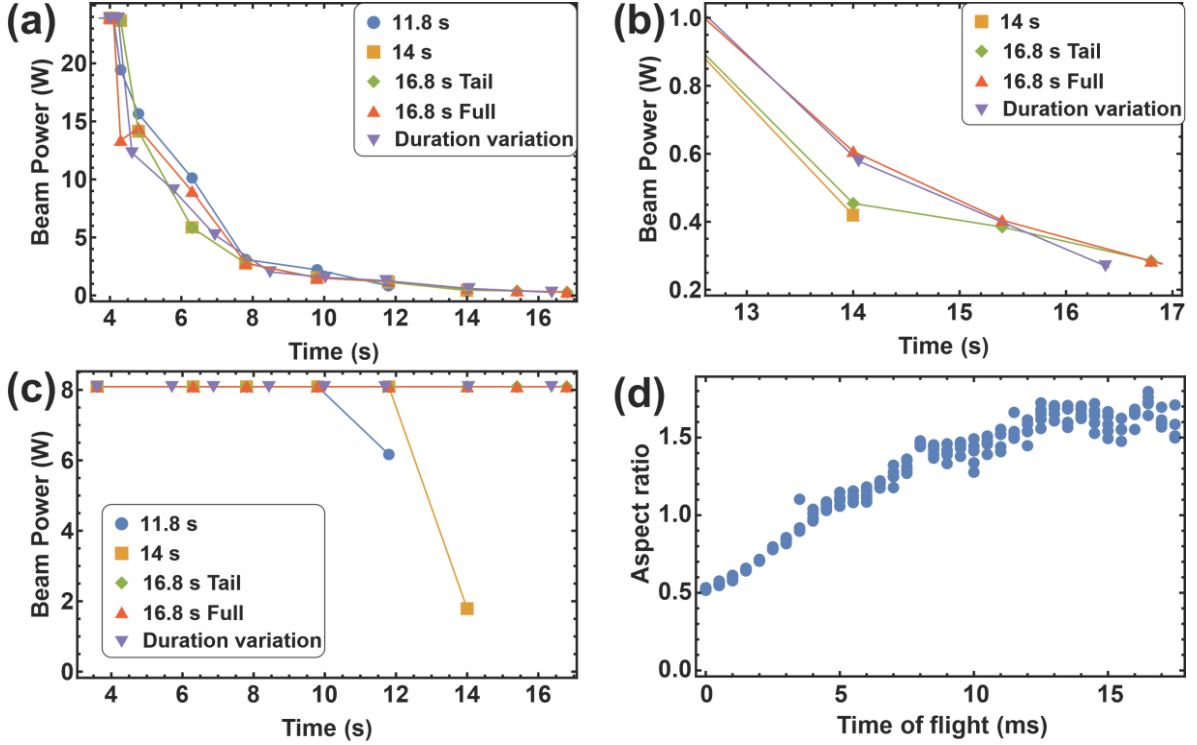

Figure 5. a) Horizontal ODT beam power in a linear scale. b) Horizontal ODT beam power in a linear scale (zoomed). c) Vertical ODT beam power. d) The aspect ratio of BEC during the free fall expansion.

## APPENDIX B MEASUREMENT OF PSD

The phase space density $D_{PS}$ of an atomic cloud was calculated as

$$D_{PS} = N\left(\frac{h\bar{v}}{k_b T}\right)^3 \qquad (3)$$

where $N$ is the number of atoms, $\bar{v}$ is the geometric mean of the $x, y, z$ trap frequencies, and $T$ is the temperature of an atomic cloud. During the evaporation, atoms from the horizontal ODT are loaded into the crossed ODT, and both of them are presented in the absorption image overlaying. The horizontal and crossed ODT clouds were separated by the double Gaussian fit, and the $D_{PS}$ value was calculated for each of them independently. That is the reason why the plots in the main text contain two separate sections.

## APPENDIX C EVAPORATION IN VARIOUS MAGNETIC FIELDS

Figure 6 shows evaporation ramps obtained by the optimization procedure with a cost function (2) (see the main text) for different magnetic fields.

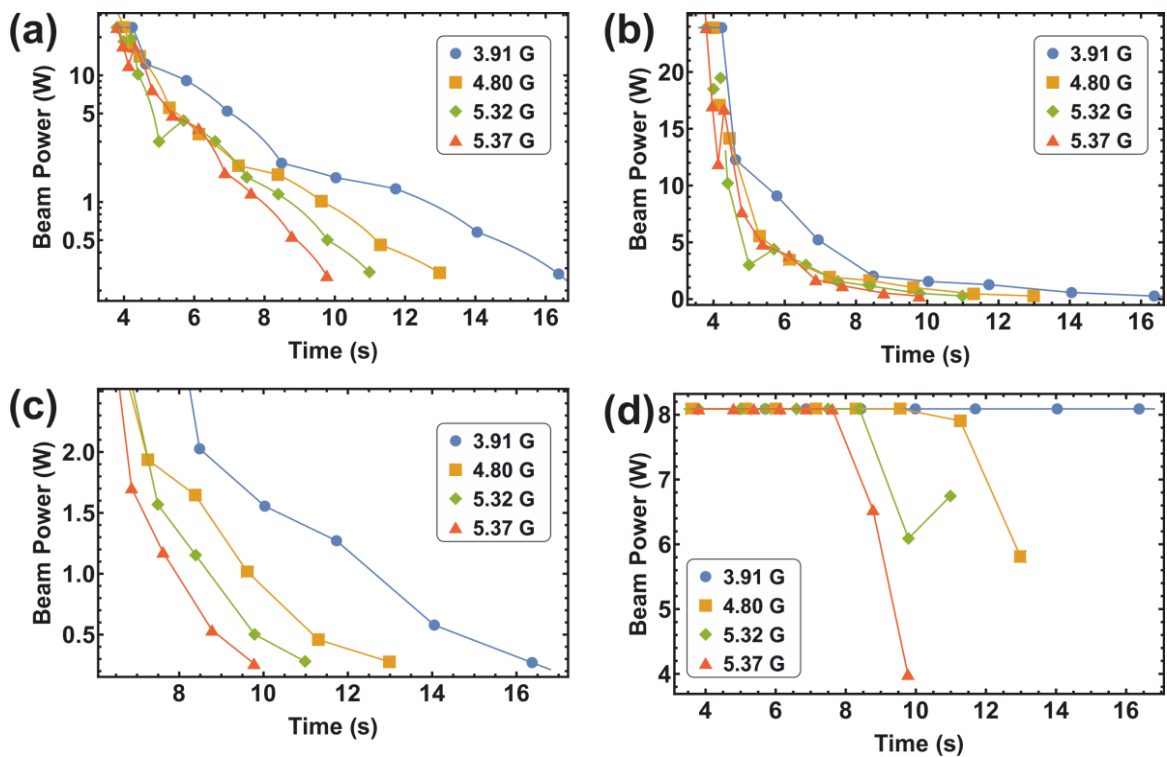

Figure 6. a) Horizontal ODT beam power in a log scale. b) Horizontal ODT beam power in a linear scale. c) Horizontal ODT beam power in a linear scale (zoomed). c) Vertical ODT beam power.